\documentclass{emulateapj}   
\usepackage{apjfonts}

\slugcomment{Accepted for publication in ApJ  (9 Mar 2007)}

\shorttitle{Infrared Study of PKS~1136$-$135 Jet}
\shortauthors{Uchiyama et al.}

\begin{document}

\title{An Infrared Study of the Large-scale Jet in Quasar PKS~1136$-$135}


\author{Yasunobu~Uchiyama,\altaffilmark{1} C.~Megan~Urry,\altaffilmark{2} 
Paolo~Coppi,\altaffilmark{2} Jeffrey~Van~Duyne,\altaffilmark{2} 
C.~C.~Cheung,\altaffilmark{3,4} 
Rita~M.~Sambruna,\altaffilmark{5} Tadayuki~Takahashi,\altaffilmark{1,6}
Fabrizio~Tavecchio,\altaffilmark{7} and Laura~Maraschi\altaffilmark{7} }

\altaffiltext{1}{Department of High Energy Astrophysics, ISAS/JAXA, 3-1-1 Yoshinodai, Sagamihara, Kanagawa, 229-8510, Japan;
uchiyama@astro.isas.jaxa.jp}
\altaffiltext{2}{Yale Center for Astronomy and Astrophysics, 
Yale University, 260 Whitney Ave., New Haven, CT 06520-8121}
\altaffiltext{3}{Jansky Postdoctoral Fellow; National Radio Astronomy
Observatory} 
\altaffiltext{4}{Kavli Institute for Particle Astrophysics and 
Cosmology, Stanford University, Stanford, CA 94305}
\altaffiltext{5}{NASA Goddard Space Flight Center, Code 661, Greenbelt,
MD 20771}
\altaffiltext{6}{Department of Physics, The University of Tokyo, 7-3-1 Hongo, Bunkyoku, Tokyo 113-0033, Japan}
\altaffiltext{7}{Osservatorio Astronomico di Brera, 
via Brera 28, 20121 Milano, Italy}

\begin{abstract}
We present \emph{Spitzer} IRAC imaging of the large-scale jet in 
the quasar PKS~1136$-$135 at wavelengths of 3.6 and $5.8\ \mu\rm m$, 
combined with previous \emph{VLA}, \emph{HST}, and \emph{Chandra} 
observations. 
We clearly detect infrared emission from the jet, resulting in 
the most detailed multifrequency data among the jets in lobe-dominated 
quasars. 
The spectral energy distributions of the jet knots 
have significant variations along the jet, like the archetypal jet in 3C~273. 
The infrared measurements with  IRAC are consistent with the previous idea that 
the jet has two spectral components, namely 
(1) the low-energy synchrotron 
spectrum extending from radio to infrared, and (2) the high-energy 
component responsible for the X-ray flux. The optical fluxes may be 
a mixture of the two components. 
We consider three radiation models for
the high-energy component: inverse Compton scattering of cosmic microwave background (CMB) photons 
by radio-emitting electrons in a highly relativistic jet, 
synchrotron radiation by a second distinct electron population, and 
synchrotron radiation by ultra high energy protons. 
Each hypothesis leads to important insights into 
and constraints on particle acceleration in the jet, as well as the basic 
physical properties of the jet such as bulk velocity, transporting power, and 
 particle contents. 
\end{abstract}


\section{Introduction}

Over the last decade, optical and X-ray observations 
made with the 
\emph{Hubble Space Telescope} \citep[e.g.,][]{Cra93,Bah95} and the \emph{Chandra} 
X-ray Observatory \citep[e.g.,][]{Cha00} have produced exquisite images 
of extragalactic kiloparsec-scale jets, completely changing our
understanding of their properties. 
Currently, more than 70 (30) extragalactic jets and hotspots are known in the X-ray (optical)
\footnote{An on-line list of extragalactic jets detected in the X-ray is found at 
\url{http://hea-www.harvard.edu/XJET/}, and that in the optical at 
\url{http://home.fnal.gov/\~{}jester/optjets}}; all but the few brightest jets 
were discovered by \emph{Chandra} (\emph{HST}).

The origin of the broad-band spectral energy distributions (SEDs)
of large-scale quasar jets, constructed 
using \emph{HST} and \emph{Chandra} data, 
are the subject of active debate \citep[for reviews, see][]{Sta03,HK06}. 
In luminous quasars with X-ray jets extending from 
the quasar nucleus out to hundreds of kiloparsec \citep[e.g.,][]{Sch00,Sam04,JM04,Mar05,Sie06}, 
the X-ray intensity relative to the radio synchrotron flux 
is generally too high to be explained by a synchrotron-self-Compton model
unless there is a huge deviation from equipartition \citep{Cha00,KS05}.
The radio, optical, and X-ray fluxes of a jet knot generally 
trace a peaked, inflected broad-band spectrum, which 
rules out the interpretation of X-rays as due to 
synchrotron radiation from a single population of 
electrons\footnote{ It should be noted that, 
unlike the case of high-power jets, 
a synchrotron interpretation of X-ray emission has been well established
for low-luminosity Fanaroff-Riley I jets \citep[e.g.,][]{Har01}.}.
An alternative scenario, 
inverse-Compton (IC) scattering of CMB photons by
high-energy electrons ($\gamma_e \sim 30$) 
in a highly relativistic jet with bulk Lorentz factor $\Gamma \sim 10$ 
\citep[the beamed IC model:][]{Tav00,CGC01} initially seemed a more natural 
way to explain the observed X-ray emission, 
but this process is also not free of problems \citep[e.g.,][]{AD04}.
Finally, it is possible that the X-rays arise from synchrotron
radiation from extremely energetic protons \citep{Aha02}. 
Determining which of these emission mechanisms produces the
observed X-ray jets in powerful quasars is a strong
motivation for more observations of radio-loud quasars,
and has resulted in a rapid increase 
in the number of known X-ray and optical jets. 

A new window to explore extragalactic large-scale jets has been opened by 
the \emph{Spitzer Space Telescope}, which is capable of 
detecting jet infrared emission thanks to the
excellent sensitivity of the 
Infrared Array Camera \citep[IRAC;][]{Faz04} and 
Multiband Imaging Photometer \citep[MIPS;][]{Rie04}.
The first example was the detection of infrared synchrotron radiation 
from jet knots in the quasar PKS~0637$-$752 
with the \emph{Spitzer} IRAC at wavelengths of 3.6 and $5.8\ \mu\rm m$ 
\citep{Uch05}. In terms of the beamed IC model, the infrared bandpass is particularly 
interesting since the bulk-Comptonization bump produced by cold electrons is expected to 
appear in the infrared \citep{Geo05}. The absence of such features in the PKS~0637$-$752 jet
 rules out the jet model dynamically dominated by $e^{+}e^{-}$ pairs 
 in the guise of the beamed IC model \citep{Uch05}. 
The MIPS observations of the jet in Centaurus A \citep{Har06,Bro06}, 
and most recently, the results from IRAC and MIPS imaging 
photometry of the jet in M~87 \citep{Shi06} have been reported, which also 
demonstrate the power of {\it Spitzer} to study jet emissions 
in lower power jets \citep[see also][for the M87 jet]{Per07}.

Now the three Great Observatories collectively offer the
possibility to identify the radiation mechanisms 
operating in powerful quasar jets.
In fact, 
when combined with the data from the \emph{VLA}, \emph{Hubble}, and \emph{Chandra} \citep{Jes05,Jes06}, 
the \emph{Spitzer} observation of the bright quasar 3C~273 
shed new light on the riddle of X-ray jets \citep{Uch06}.
The \emph{Spitzer} IRAC photometry of the jet knots in 3C~273 
indicated a two-component spectral energy distribution:
a radio-to-infrared synchrotron component and a 
separate optical-to-X-ray component.
The latter also seems likely to be of synchrotron origin, 
given the similar polarization of optical and radio light. 
The optical polarization, however,  has not yet been measured 
with high precision, so this conclusion is not yet firm.
Perhaps such a double synchrotron scenario is applicable 
to the radiation output from many quasar jets. 

In this paper, we present \emph{Spitzer} IRAC imaging 
of the powerful jet in the luminous quasar PKS~1136$-$135. 
Together with data from the \emph{VLA}, \emph{HST}, and \emph{Chandra} 
\citep{Sam02,Sam04,Sam06a}, our infrared
photometry makes the SED of the PKS~1136$-$135 jet the most detailed 
and best constrained among {\it lobe-dominated} quasars. 
The jet in the quasar PKS~1136$-$135 is 
reminiscent of the 3C~273 jet \citep{Sam06a}, 
demonstrating anti-correlation between radio and X-ray brightness,
such that the radio intensity increases toward a hotspot 
while X-ray flux decreases. 
Applying the beamed IC model to the X-ray emission, this has recently been 
interpreted to imply {\it deceleration} of the jet \citep{Sam06a,Tav06}.
Here we analyze the multiwavelength jet emission in the light of 
the double synchrotron scenario recently outlined 
for the 3C~273 jet \citep{Uch06}.
The redshift of PKS~1136$-$135 is $z=0.554$, so we 
adopt a luminosity distance of $D_L = 3.20\ \rm Gpc$, 
for a concordance cosmology with
$\Omega_{\rm m}=0.27$, 
$\Omega_{\Lambda}=0.73$, and 
$H_0=71\ \rm km\ s^{-1}\ Mpc^{-1}$.
The angular scale of $1\arcsec$ corresponds to 6.4 kpc.

\section{Observations and Results}
\label{observation}

\subsection{Spitzer IRAC observations}
\label{irac}

\begin{figure} 
\epsscale{1.2}
\plotone{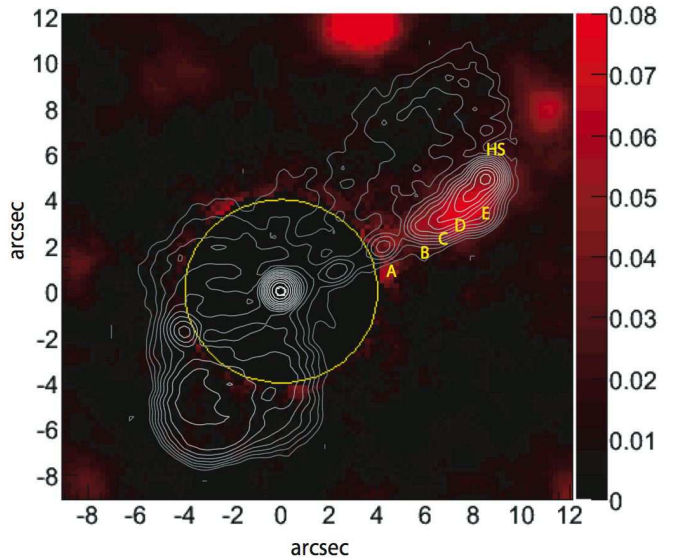}
\caption{IRAC image of the PKS~1136$-$135 jet 
at $3.6\ \mu\rm m$ in units of $\rm MJy\ sr^{-1}$ with core subtraction. 
North is up and east is to the left. The origin of the coordinate is set 
at the quasar core. A circular region with a radius of $4\arcsec$ around the core 
is blanked out, where the core subtraction is incomplete. 
An angular scale $1\arcsec$ corresponds to 6.4 kpc projected size. 
Superimposed on the IRAC image are 
radio contours on a logarithmic scale from the VLA 8.5 GHz map. \label{fig:image} }
\end{figure}

We observed 
 PKS~1136$-$135 with \emph{Spitzer} IRAC \citep{Faz04} 
on 2005 June 10 as part of our Cycle-1 General Observer program 
(\emph{Spitzer} program ID 3586). 
We used the pair of 3.6 and $5.8\ \mu\rm m$ arrays, observing 
the same sky simultaneously. 
The pixel size in both arrays is $\simeq 1\farcs22$. The point-spread 
functions (PSFs) are $1\farcs66$ and $1\farcs88$ (FWHM) for the 
3.6 and $5.8\ \mu\rm m$ bands, respectively.
The photometry with IRAC is calibrated to an accuracy of 
$\sim 3\%$ \citep{Rea05}.
We obtained a total of 50 frames per IRAC band, each with a 30-s frame time.
The pipeline process (version S14.0.0) at the \emph{Spitzer} Science Center 
yielded 50 calibrated images (Basic Calibrated Data). 
These well-dithered frames were combined into a mosaic image with 
a pixel size of $0\farcs2$ using MOPEX \citep{MM05}, 
which removes spurious sources such as cosmic rays and moving objects based on inter-frame comparisons. 
Finally, in order to align the IRAC image with respect to the VLA image, 
we set the center of the quasar in the IRAC image at the 
core position in the VLA image. In this way IRAC astrometry 
is estimated to be accurate to $\la 0\farcs2$. 

The infrared fluxes from the quasar core of PKS~1136$-$135 were measured as 
2.2 and 4.0 mJy in the 3.6 and $5.8\ \mu\rm m$ bands, respectively
(Table~\ref{tbl-1}).
These values do not reach the saturation limits of the IRAC arrays. 
Assuming that the infrared spectrum has a power law with $\alpha =1$, 
as consistent with the ratio of the 3.6 and $5.8\ \mu\rm m$ fluxes, 
the luminosity in the 3--10 $\mu\rm m$ band amounts to 
$L_{3\mbox{--}10 \mu \rm m} \simeq 2.6 \times 10^{45}\ \rm ergs\ s^{-1}$.
In the model of \citet{Sam06b}, 
the infrared emission is assumed to arise from the inner jet, while the optical/UV 
emission originates in the accretion disk. Our infrared measurement is consistent 
with that model.
If this is the case, the infrared fluxes determine 
the apparent luminosity of the synchrotron component. 

Inspection of the $3.6\-\mu\rm m$ IRAC images 
clearly reveals the presence of an infrared jet that traces 
the extended radio jet.
The IRAC $5.8\ \mu\rm m$ image also shows the jet 
emission, although its quality is worse than the 
$3.6\ \mu\rm m$ image. 
The extended PSF of the bright quasar core is significant at the 
location of the jet, 
amounting to $\sim 10\mbox{--}30\%$ of the jet flux. 
In order to separate the jet infrared emission from the contaminating PSF 
wings of the core, we subtracted the PSF wings 
 by making use of the PSF templates in a way similar to \citet{Uch05}.
Figure \ref{fig:image} shows the IRAC image of PKS~1136$-$135 in 
the $3.6\ \mu\rm m$ band 
after subtraction of the PSF wings of the quasar core.
Infrared counterparts of the jet knots 
are clearly visible in the IRAC image, 
located $\sim 8\arcsec \mbox{--}12\arcsec$ from 
the core (knot B through hotspot HS).
A possible counterpart of knot A is marginally found in the IRAC image, 
but it may be due to our incomplete PSF subtraction, so we regard 
the knot A flux only as an upper limit. 
The hotspot at the counter-jet side also suffers from the uncertainties of 
PSF subtraction; we do not find an infrared counterpart and place an upper limit of 
 $f_{3.6} = 4\ \mu\rm Jy$ at  $3.6\-\mu\rm m$.
Also, we do not detect extended infrared emissions from the radio lobes.

To construct the broad-band SEDs of jet knots
using multifrequency data, we 
derived infrared flux densities of some knot features at 3.6 and $5.8\ \mu\rm m$. 
Given the small separation of adjacent knots, typically $\sim 1\arcsec$, 
we utilized  a PSF-fitting method \citep{Uch06}. 
Specifically, the jet image was fitted with 
a series of IRAC PSFs, fixing 
the PSF centers at the 
knot positions in the VLA 22 GHz radio image. 
However, this procedure did not fully determine knot by knot fluxes 
due mainly to a multiplicity of knots from C to E. 
We then fixed the flux density of knot E at 
$f_{3.6} = 2\ \mu\rm Jy$ and $f_{5.8} = 4\ \mu\rm Jy$. 
We derived the systematic errors due to this treatment by changing 
the knot E flux in a range of $f_{3.6} = 0\mbox{--}3\ \mu\rm Jy$ 
and $f_{5.8} = 0\mbox{--}6\ \mu\rm Jy$. 
The systematic errors are then added in quadrature to statistical ones. 
The photometric results determined in this way 
are listed in Table~\ref{tbl-1}, where 
a  combined flux is reported for knots C, D, and E (referred as knots CDE).

\begin{deluxetable*}{crrrrrrrrr}
\tablecolumns{10}
\tabletypesize{\footnotesize}
\tablecaption{Flux Densities of the PKS~1136$-$135 Core and Its Jet Knots\label{tbl-1}}
\tablewidth{0pt}
\tablehead{
\multicolumn{1}{c}{} &
\multicolumn{9}{c}{Flux Density, $f_{\nu}$ }\\
\cline{2-10}\\ 
\colhead{Region} 
& \colhead{$f_{5\ \rm GHz}$} & \colhead{$f_{8.5\ \rm GHz}$} 
& \colhead{$f_{22\ \rm GHz}$} & \colhead{$f_{5.8\ \rm \mu m}$}
& \colhead{$f_{3.6\ \rm \mu m}$} & \colhead{$f_{0.81\ \rm \mu m}$}
& \colhead{$f_{0.48\ \rm \mu m}$} 
& \colhead{$f_{1.56\ \rm keV}$} & \colhead{$f_{4.35\ \rm keV}$} \\
\colhead{} & \colhead{(mJy)} & \colhead{(mJy)} & \colhead{(mJy)} & 
\colhead{($\mu$Jy)} & \colhead{($\mu$Jy)} & \colhead{($\mu$Jy)} &
\colhead{($\mu$Jy)} & \colhead{(nJy)} & \colhead{(nJy)} \\
}
\startdata
Core 
& 463 & \nodata & \nodata 
& $4.0\times 10^3$ & $2.2\times 10^3$ & \nodata & $1.5\times 10^3$ 
& 170 & \nodata \\
A 
& \nodata & $3.8\pm 0.4$ & \nodata 
& $<5$ & $<4$ & $0.23\pm 0.07$ & $0.17\pm 0.03$ 
& $0.77\pm 0.22$ & $0.72\pm 0.29$ \\
B 
& $11\pm 1.6$ & $9.3\pm 0.9$ & $3.2\pm 0.6$ 
& $4.8 \pm 2.2$ & $3.8 \pm 1.9$ & $0.33\pm 0.09$ & $0.22\pm 0.03$ 
& $2.0\pm 0.34$ & $0.72\pm 0.24$ \\
CDE
& $192\pm 19$ & $116\pm 12$ & $47.5\pm 9.5$ 
& $20\pm 4$ & $9.5\pm 2.0$ & $<0.46$ & $<0.23$ 
& $1.2\pm 0.26$ & $0.67\pm 0.25$ \\
HS 
& $200\pm 20$ & $119\pm 12$ & $43.9\pm 8.8$ 
& $7.4\pm 2.7$ & $4.6\pm 2.0 $ & $0.24\pm 0.05$ & $<0.08$ 
& $<0.16$ & \nodata \\
\enddata
\tablecomments{ 
The radio and optical data (extinction-corrected) for jet knots are 
taken from \citet{Sam06a}. 
The errors of optical 
fluxes are translated into those at a 90\% level. The infrared fluxes of 
the jet knots are determined through 
PSF-fitting and quoted errors are dominated by their systematic errors. 
The X-ray fluxes listed with statistical errors  (at a 90\% confident level) 
are corrected for Galactic absorption using 
$N_{\rm H} = 3.5\times 10^{20}\ \rm cm^{-2}$. Except for the infrared, the data of the core fluxes are taken from \citet{Gam03} assuming $\alpha_{\rm opt} = 0.5$ and 
$\alpha_{\rm X} = 0.7$.}
\end{deluxetable*}

\subsection{Multifrequency Image}

\begin{figure} 
\plotone{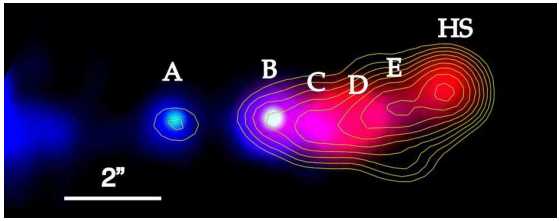}
\caption{Composite image of the jet in PKS~1136$-$135 based on 
the data from \emph{Spitzer} (shown as {\it red}), \emph{HST} ({\it green dots} 
on knots A and B), 
and \emph{Chandra} ({\it blue}). 
 The superposed contours are from the \emph{VLA} (8.5 GHz), highlighting the 
 jet radio-emission (see also Fig.\ \ref{fig:image}).
Letters label the main features along the jet; knot A closest to the quasar.
The multiwavelength image qualitatively illustrates the relative importance of 
low-energy component (radio--infrared) in the outer knots
and high-energy one (optical--X-ray) in the inner knots.
\label{fig:great} }
\end{figure}

In Figure \ref{fig:great} 
we present a three-color multifrequency image based on 
\emph{Spitzer} infrared ({\it red}), \emph{HST} optical ({\it green}), 
and \emph{Chandra} X-rays ({\it blue}). 
The VLA 8.5 GHz radio contours in a logarithmic scale are overlaid.
The infrared photometry at $3.6\ \mu\rm m$ is 
illustrated as a series of best-fitted PSFs of every knot after 
artificially shrinking the widths of the PSFs ($1\arcsec$ FWHM) to 
restore a resolution similar to the X-rays.
At the positions of optically bright knots A and B we plot a {\it green} dot of $\sigma \simeq 0\farcs 2$ to illustrate clear detections with the \emph{HST} STIS and ACS 
\citep{Sam04,Sam06a}.
The \emph{Chandra} X-ray  image in 0.3--8 keV  band is 
smoothed with a Gaussian kernel of $0\farcs 4$.

The jet morphology, or brightness pattern along the jet, in the infrared appears 
to be different from emission at shorter wavelengths. 
It is interesting to note that optically-bright features (knots A and B)
are not particularly prominent in the infrared, while an infrared-bright knot D
does not emit correspondingly strong optical light. 
The difference in the brightness pattern between the infrared and 
optical suggests
a dramatic change of the spectral shape in the infrared-to-optical band 
along the jet. This emphasizes the importance of measuring infrared 
fluxes, together with the optical fluxes, in quasar jets. 

Putting hotspot HS aside, the jet knots can be divided up into 
two parts: 
the {\it inner knots} (A and B) and the {\it outer knots} (C, D, and E).
The inner knots are bright in both the optical and X-rays, and as such 
they are ``high-energy dominated." 
Indeed the inner knots are most luminous in the X-rays, while 
the outer knots seem to have a luminosity peak in the far-infrared. 
To clarify this, we present the SEDs of the jet knots 
in the following section.

\subsection{Spectral Energy Distributions: Two-Component Nature}
\label{sec:SED}

\begin{table}
\begin{center}
\caption{Phenomenological Model Parameters of PKS~1136$-$135 Jet \label{tbl-2}}
\begin{tabular}{ccccc}
\tableline\tableline
Parameter & knot A & knot B & knots CDE & hotspot HS \\
\tableline 
$\alpha_1$ & 0.65 & 0.65 & 0.85 & 1.0 \\
$\nu_{1}$ (Hz) & 
$<3\times10^{13}$ & $2\times10^{13}$ & $3\times10^{13}$ & $5\times10^{13}$ \\
$L_1$ ($\rm erg\ s^{-1}$) & 
$<1.3\times10^{43}$ & $3.1\times10^{43}$ & $1.7\times10^{44}$ & $1.2\times10^{44}$ \\
$\alpha_2$ & 0.75 & 0.70 & 0.50 & \nodata \\
$L_2$ ($\rm erg\ s^{-1}$) & 
$5.0\times10^{43}$ & $8.0\times10^{43}$ & $7.3\times10^{43}$ & \nodata \\
\tableline
\end{tabular}
\tablecomments{Equation (\ref{eq:fnu}) is adopted to characterize the SEDs.}
\end{center}
\end{table}

We construct the broad-band SEDs of the jet knots
using the new \emph{Spitzer} IRAC data added to the 
\emph{VLA}, \emph{HST}, and \emph{Chandra} data
already presented in \citet{Sam06a}. 
Once again, the \emph{Spitzer} IRAC provides 
crucial photometric points in characterizing the emissions 
of quasar jets \citep{Uch05,Uch06}.

Table \ref{tbl-1} summarizes the flux densities obtained with 
\emph{VLA}, \emph{Spitzer}, \emph{HST}, and \emph{Chandra}. 
The radio fluxes  measured with the VLA at frequencies of 4.9, 8.46, 
and 22.5 GHz are taken from \citet{Sam06a}. 
Infrared photometry with the \emph{Spitzer} IRAC is described in \S\ref{irac}. 
In the optical, we list the flux densities from the \emph{HST} ACS 
 in two filters \citep{Sam06a}:
 F475W ($0.48\ \mu\rm m$) and F814W ($0.81\ \mu\rm m$). 
We employed a somewhat different method for 
X-ray photometry  compared with \citet{Sam06a}.
Specifically, we here define two energy bands for photometry:  1--2 keV and 3--6 keV.
Integrating over the entire jet (from knot A to hotspot HS) we found 203 and 57 counts 
in the 1--2 and 3--6 bands, respectively (also, 147 counts in the 0.6--1 keV band).
The nominal energy $\varepsilon_{\rm norm}$ for each band is 
the mean energy weighted by the effective area: $\varepsilon_{\rm norm}=$ 
1.56 keV and 4.35 keV. 

We reanalyzed the {\it  Chandra} data sets that are fully described in \citet{Sam06a}, 
and constructed flux images (in units of $\rm photons\ cm^{-2}\, s^{-1}\, pix^{-1}$) 
in the two bands, 
to which aperture photometry was applied.
The object apertures for knots A/B and hotspot are 
a  circle with a radius of $\simeq 0\farcs 7$ centered on each feature, while 
 that for knots CDE is a box enclosing knots C, D, and E. 
To correct for the interstellar absorption of 
$N_{\rm H} = 3.5\times 10^{20}\ \rm cm^{-2}$, 
we multiplied the measured fluxes by a  factor of 1.04
for the 1--2 keV band. The fudge factor slightly depends on 
the source spectrum such that it ranges from 1.040 to 1.048 for 
a power law with photon index  $\Gamma = 1\mbox{--}3$.
The photometry-based spectra are found to be consistent with 
the usual spectral analysis presented by \citet{Sam06a}.
The photometric method is convenient when combining lower frequency data.

Care should be taken in interpreting the X-ray fluxes. 
First, the X-ray emission downstream of knot B does not show 
well discernible knots, and therefore the X-ray flux of knots CDE 
may not be related directly to the lower frequency fluxes. 
Second, 
in knot A there is an offset of $\sim 0\farcs 4$ between
the soft and hard X-ray peaks \citep{Sam06a}. 
The X-ray spectrum derived for knot A may be contaminated by 
some unrelated emission, which makes the spectral shape inconsistent 
with a power law.

\begin{figure}
\plotone{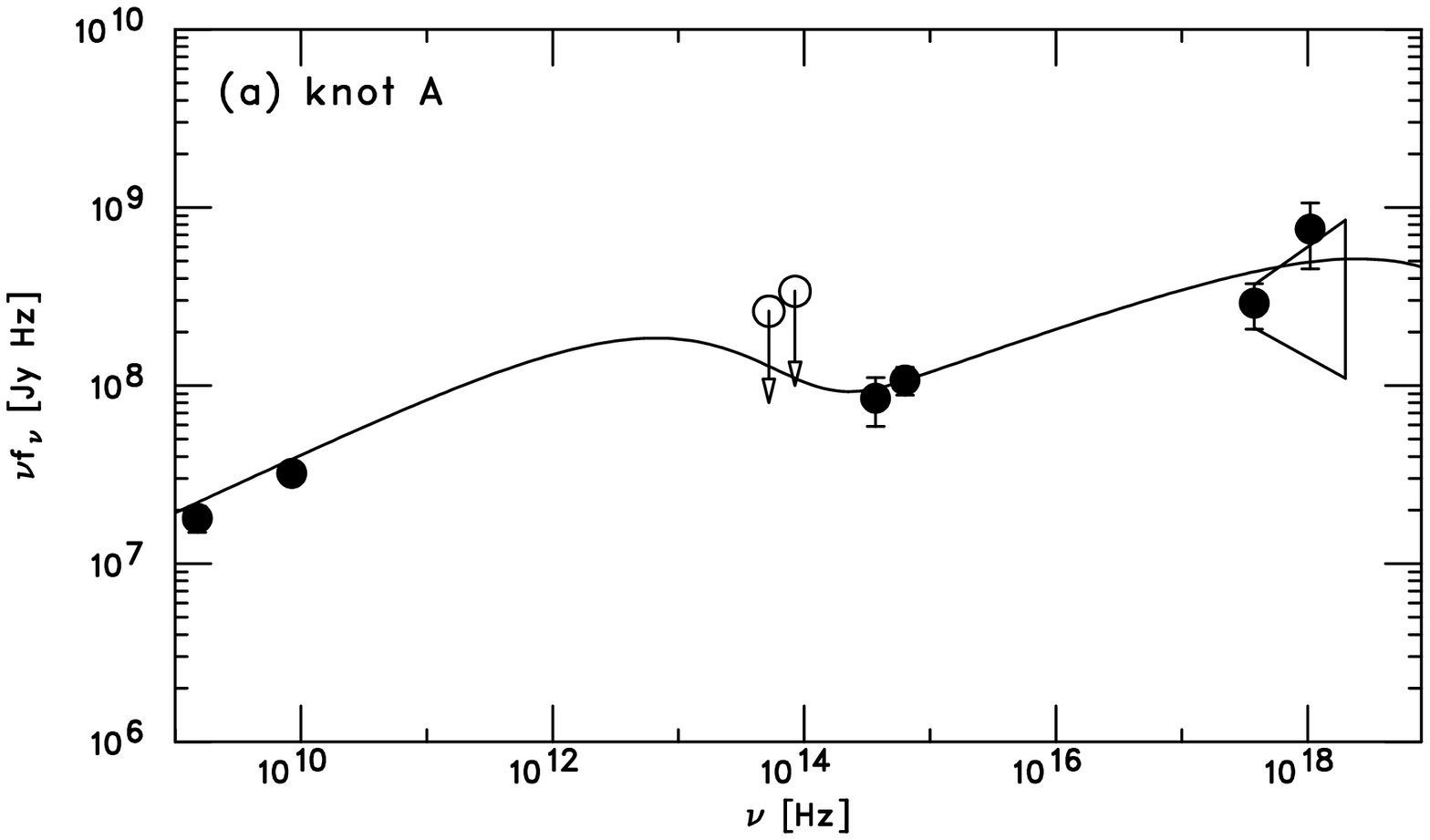}
\plotone{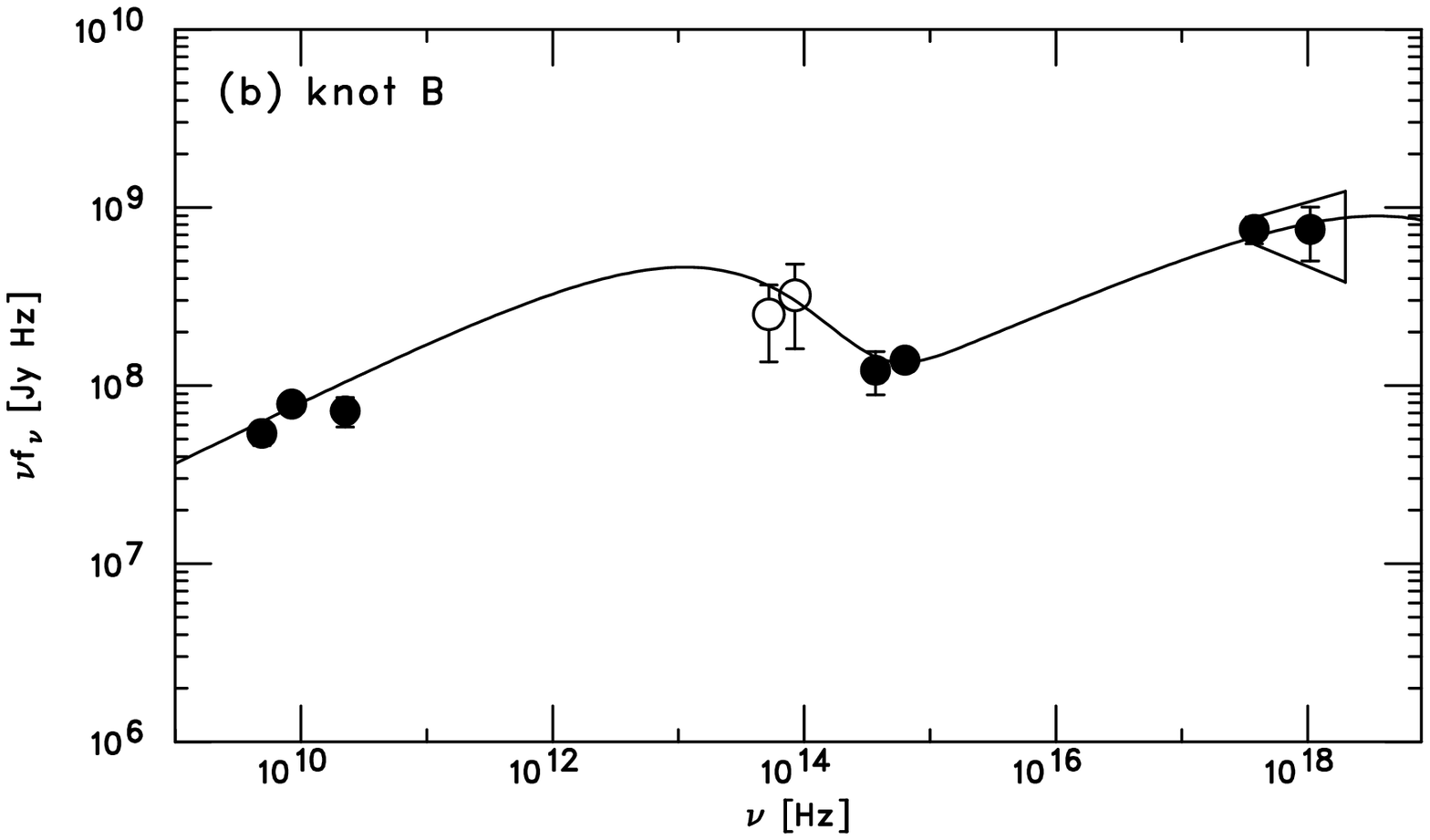}
\plotone{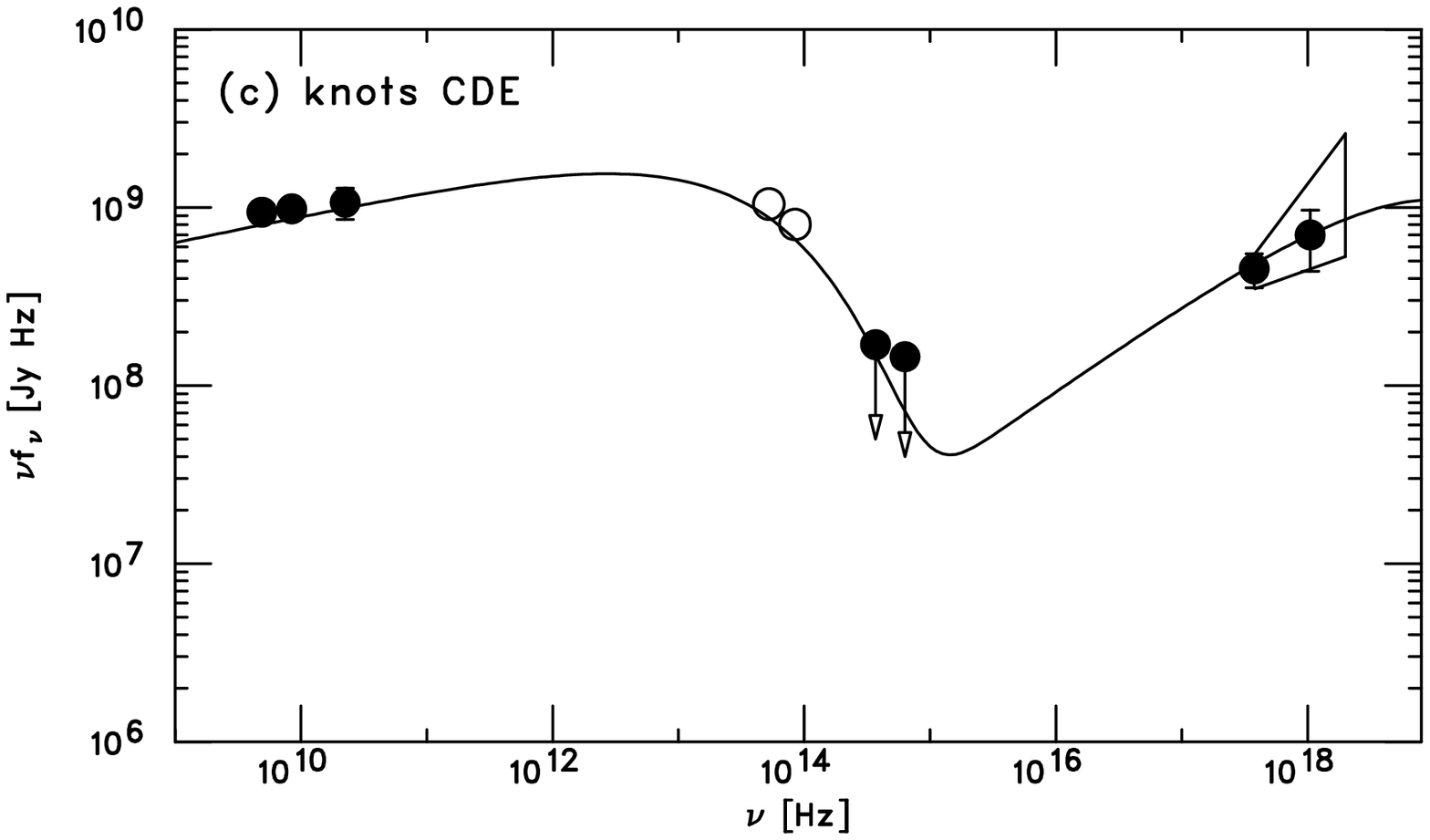}
\plotone{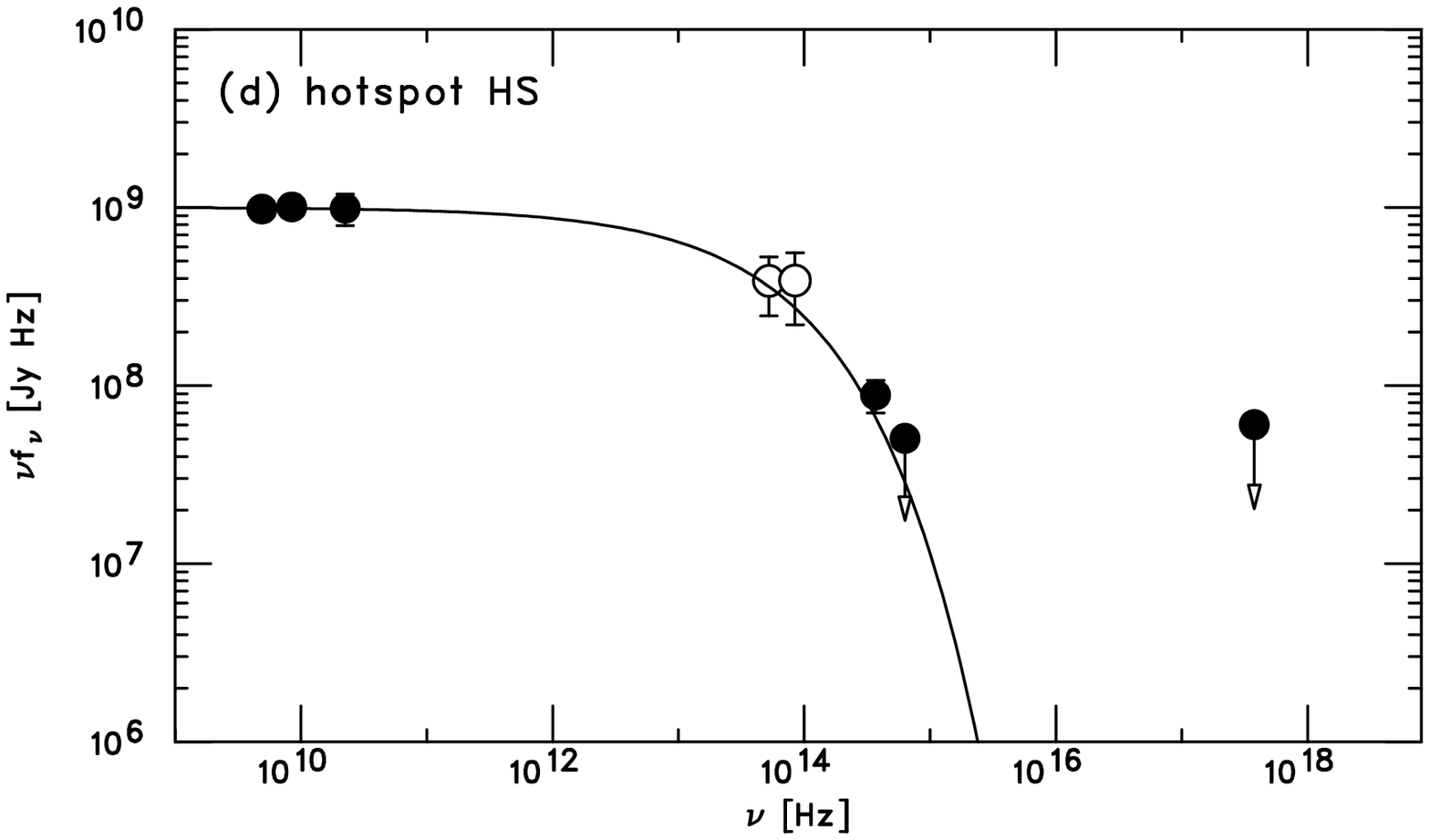}
\caption{Broadband radio-to-X-ray SEDs for knots A, B, CDE, and hostspot HS
in the powerful quasar 1136$-$135. IRAC 3.6 and 
$5.8\ \mu\rm m$ fluxes are shown 
as open circles. The data are summarized in Table \ref{tbl-1}. In addition, 
a flux density of $12.1\pm 1.8$ mJy at 1.5 GHz \citep{Sam06a} is plotted in the SED of knot A.
The two-component phenomenological model given as equation (\ref{eq:fnu}) 
is used to describe the data. 
\label{fig:SED}}
\end{figure}

In Fig.~\ref{fig:SED}, we present the SEDs from radio to X-rays for 
knots A, B, CDE, and hotspot HS. 
We note that the infrared fluxes we have measured with the \emph{Spitzer} 
IRAC fill central points in the radio-to-X-ray SEDs 
of the jet knots, setting an important constraint on models 
of the broad-band emission.
The {\it inner knots} (A and B) radiate most strongly 
in the X-ray.
Also, significant (and probably flatter) optical emission is observed 
exclusively from the {\it inner knots}.
The possible difference in slopes of the optical continuum between 
the {\it inner} and {\it outer knots} \citep{Sam06a} suggest different origins; 
the optical emission in the {\it inner knots} may belong to the same spectral 
component responsible for the X-ray emission \citep[as suggested by][for knot A]{Sam06a},
while in the outer knots the optical emission 
is related to the radio synchrotron component.

These spectral characteristics are analogous to those of 
the jet in 3C~273 \citep{Uch06}, for which we identified 
two spectral components: (1) the low-energy synchrotron 
spectrum extending from radio to infrared, and (2) the high-energy 
component arising in the optical and smoothly connecting to the X-ray flux. 
We argued that the second component is likely to be of synchrotron 
origin as well, thus forming double-synchrotron spectra, 
because of the similarity of polarization between optical and radio
emission (though the degrees of optical polarization are not established yet).

Given the similarity to 3C~273, 
we model the radio-to-X-ray SEDs phenomenologically 
by the following function, namely, a double power-law with an exponential cutoff: 
\begin{equation}
\label{eq:fnu}
f_{\nu} =
\sum_{i=1}^{2}\kappa_i \nu^{-\alpha_i}
\exp\left[ -\left( \frac{\nu}{\nu_{i}} \right)^{1/2}\right] .
\end{equation}
The first term ($i=1$) of the right-hand-side accounts for the low-energy spectrum and 
the second term ($i=2$) describes the high-energy part.
As there are no firm indication of spectral steepening at X-rays, we set an arbitrary 
cutoff at $\nu_2 = 1\times 10^{19}$ Hz for the second component.
For the inverse-Compton model described later (\S\ref{sec:ICCMB}; 
see Fig.~\ref{fig:ICCMB}), however, 
the high-energy cutoff is located far beyond the X-ray domain,  reaching 
very high-energy gamma-rays.

Table \ref{tbl-2} lists the spectral parameters that well 
describe the knot SEDs shown in Fig.~\ref{fig:SED}.
Instead of the normalization, $\kappa_i$, we present the total apparent luminosity 
(without taking account of relativistic beaming) 
defined as $L = 4\pi D_L^2 \int f_\nu d\nu$ for each component, 
where the low-energy boundary of the integral is set 
at $10^{9}$ Hz and $10^{13}$ Hz for the first and second 
component, respectively. 
Knot B, the most X-ray luminous one, has a luminosity of 
$L_2 \simeq 8\times 10^{43}\ \rm erg\ s^{-1}$ 
for the high-energy component, 
which indeed exceeds the luminosity of the low-energy component. 

The spectral index of the low-energy component is in the range of 
 $\alpha_1 \simeq 0.65\mbox{--}1.0$, which is 
 determined by the radio spectra, as listed in Table \ref{tbl-2}.
The cutoff frequency of the low-energy component, as constrained 
by the IRAC fluxes, is 
$\nu_1 = (2\mbox{--}5)\times 10^{13}$ Hz in the cases of knots B, CDE, and 
hotspot HS; $\nu_1 < 3 \times 10^{13}$ Hz for knot A. 
The second spectral index is found to be $\alpha_2 \simeq 0.5\mbox{--}0.75$.

\section{Interpreting the High-Energy Component}

While the low-energy emission from quasar jets
is undoubtedly of synchrotron origin, 
the production mechanism(s) of strong X-rays 
remains a matter of debate \citep[e.g.,][]{HK06}.
Three radiation models have been proposed 
to explain the high-energy component: 
inverse-Compton scattering by radio-emitting electrons, 
synchrotron radiation by a second electron population, 
and synchrotron radiation by energetic protons\footnote{Yet another possibility includes the model of 
\citet{DA02}, which invokes spectral hardening due to inverse-Compton cooling in 
the Klein-Nishina regime.}.
In this section we consider each of these in turn. 

There are reasons to think that jets are relativistic on large scales \citep{Gar88} 
and thus beaming may be important.
The Doppler beaming factor is defined as $\delta \equiv [\Gamma(1-\beta\cos\theta)]^{-1}$ where $\beta c$ is the velocity of the jet, 
 $\Gamma = (1-\beta^2)^{-1/2}$ is the bulk Lorentz factor of the jet, 
and $\theta$ is the observing angle with respect to the jet direction.
A one-sided jet with two-sided hotspots in this source is a sign of substantial 
 relativistic beaming. 
The Doppler factor is, however, not well constrained for this jet. 
Throughout this section, all physical quantities are referred to a jet co-moving 
frame unless otherwise specified. 
The exceptions to this are the direction and velocity of the jet itself. 
It is assumed that a relativistically moving blob (as noted by knot or hotspot) has 
spherical geometry with a radius of $r=1$ kpc (in the jet frame of reference)\footnote{
The size of the emission region is not certain, as 
the knots are unresolved. The knot size is assumed to be similar to the 3C 273 jet.}
occupied homogeneously with relativistic particles and magnetic fields
(``one-zone" model with a filling factor of unity).

Before we discuss the high-energy emission component of the {\it jet knots}
(like knot B), we derive a lower limit on the magnetic field strength in  hotspot HS, 
based on the {\it absence} of the high-energy emission. 
We do not introduce beaming effects for hotspot HS ($\Gamma=\delta=1$).
Synchrotron photons are Compton upscattered by synchrotron-emitting electrons 
themselves,  namely  ``synchrotron self-Compton" (SSC),
which is known to be a primary component for the high-energy emission in 
bright hotspots such as those in Cygnus A \citep{HCP94}.
In fact, 
since the average energy density of synchrotron 
photons \citep[see][]{BG85} in hotspot HS, 
$U_{\rm syn} \simeq 9L_{\rm syn}/(16\pi r^2 c) \simeq 100\ \rm eV\ cm^{-3}$, 
largely exceeds the CMB energy density of 
$U_{\rm CMB} = 0.26(1+z)^4\ \rm eV\ cm^{-3} \simeq 1.5\ \rm eV\ cm^{-3}$, 
the SSC flux well exceeds the IC/CMB flux in the hotspot of 
PKS~1136$-$135.
We found a lower limit on magnetic field as $B >  0.2\ \rm mG$ 
in order not to violate the X-ray upper limit by the SSC emission. 
With an equipartition magnetic filed of $B_{\rm eq} = 0.66\ \rm mG$ 
where the energy density of relativistic electrons is equal to that of 
magnetic fields,  the one-zone SSC flux is an order of magnitude lower than the 
X-ray upper limit. 

\subsection{Inverse-Compton Scattering by Ultra-Relativistic Jet}
\label{sec:ICCMB}

We first discuss 
the second, X-ray-dominated component of the SED seen in the jet knots 
(knot B, in particular) in the framework of the beamed inverse-Compton model 
\citep{Tav00,CGC01}, which has been advocated in the previous 
papers for this jet \citep{Sam06a,Tav06}.
In this model, it is assumed that 
the jet has a highly relativistic bulk velocity, 
with a Lorentz factor 
$\Gamma \sim 10$ out to distances of hundreds of kiloparsecs, so that 
the CMB field {\it seen} by electrons is enhanced 
by a factor of $\Gamma^2$ in the jet comoving frame.
The amplified CMB 
can be Compton up-scattered by 
high-energy electrons of $\gamma_{e} \sim 30$ into the observed X-rays.
This model automatically requires that the jet direction is close to the 
line-of-sight, $\theta \sim \Gamma^{-1}$.
(Such a condition, and therefore the beamed IC model itself, would not 
be favorable for a lobe-dominated quasar like PKS~1136$-$135. Later in this section, we shall briefly discuss this issue.)

It is interesting to test the beamed IC/CMB hypothesis, 
particularly for the PKS~1136$-$135 jet.
The jet exhibits 
the evolution of multiwavelength emission along the jet, namely 
the increase of 
radio brightness accompanied by the decrease of X-ray 
brightness towards downstream. 
As detailed  in \citet{GK04}, such a behavior can be 
interpreted, 
within the framework of the beamed IC radiation, 
as being due to deceleration of the jet \citep{Sam06a,Tav06}.
Specifically, \citet{Tav06} concluded that 
the bulk Lorentz factor of the jet decreases from 
$\Gamma \simeq  7$ (knot B) to $\Gamma \simeq  2$ (knot E).
If the beamed IC scenario is confirmed, we gain a new probe of 
the flow structure of the large-scale jets. 

Let us apply a synchrotron plus beamed IC/CMB model of \citet{Tav00} to 
 the broadband SED of knot B.
We also include SSC calculation for completeness. 
The both models  adopt ``one-zone" emission, 
assuming  an emitting blob with homogeneously filled with relativistic particles and magnetic fields.
It is interesting to note that in the case of no-beaming ($\Gamma=\delta=1$),
again, the average energy density of synchrotron photons in the knot,
$U_{\rm syn} \simeq 12\ \rm eV\ cm^{-3}$, 
exceeds the CMB energy density.
The energy distribution of radiating electrons 
 is assumed to be of the form 
$N(\gamma ) = k \gamma^{-s}$, having  a  low-energy cutoff 
at $\gamma_{\min}=20$ and a high-energy exponential cutoff at $\gamma_{\max}$, 
where $\gamma$ denotes the electron's  Lorentz factor.  
The number index of electrons is set to be  $s=2\alpha +1 = 2.4$ based on 
a typical radio index in this jet ($\alpha \sim 0.7$), and 
the high-energy  cutoff is determined to be 
$\gamma_{\max}=4\times 10^5$ to produce the infrared flux. 
The magnetic field strength is adopted as $B=B_{\rm eq, \delta=1}\delta^{-1}$, 
where $B_{\rm eq, \delta=1}=175\,\mu$G; this relation ensures 
equipartition between relativistic electrons 
and magnetic fields, and keeps 
  the peak frequency of the synchrotron radiation, 
$\nu_{\max} \propto \delta B \gamma_{\max}^2$.
The bulk Lorentz factor itself does not explicitly affect the discussion made here 
and it is adopted to be $\Gamma = 15$.

\begin{figure}
\epsscale{1.1}
\plotone{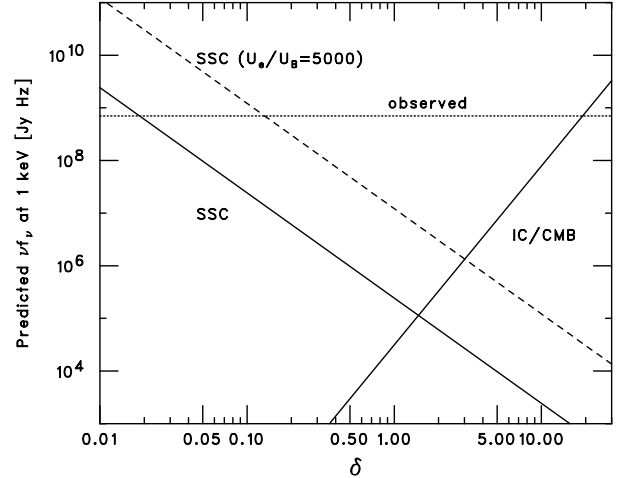}
\caption{Predicted flux of IC/CMB and SSC at 1 keV 
as a function of the beaming factor ($\delta$) based on the observed radio flux 
of knot B. The observed X-ray flux is $7\times 10^8\ \rm Jy\, Hz$ (a horizontal 
{\it dotted} line). The IC/CMB emission is dominant for a large $\delta$, while SSC dominates the X-ray emission for de-beaming cases ($\delta <1$). A dashed line shows 
the SSC prediction with the assumption of a significant deviation 
from equipartition ($U_e/U_B = 5000$).
 \label{fig:ICSSC}}
\end{figure}

\begin{figure}
\epsscale{1.15}
\plotone{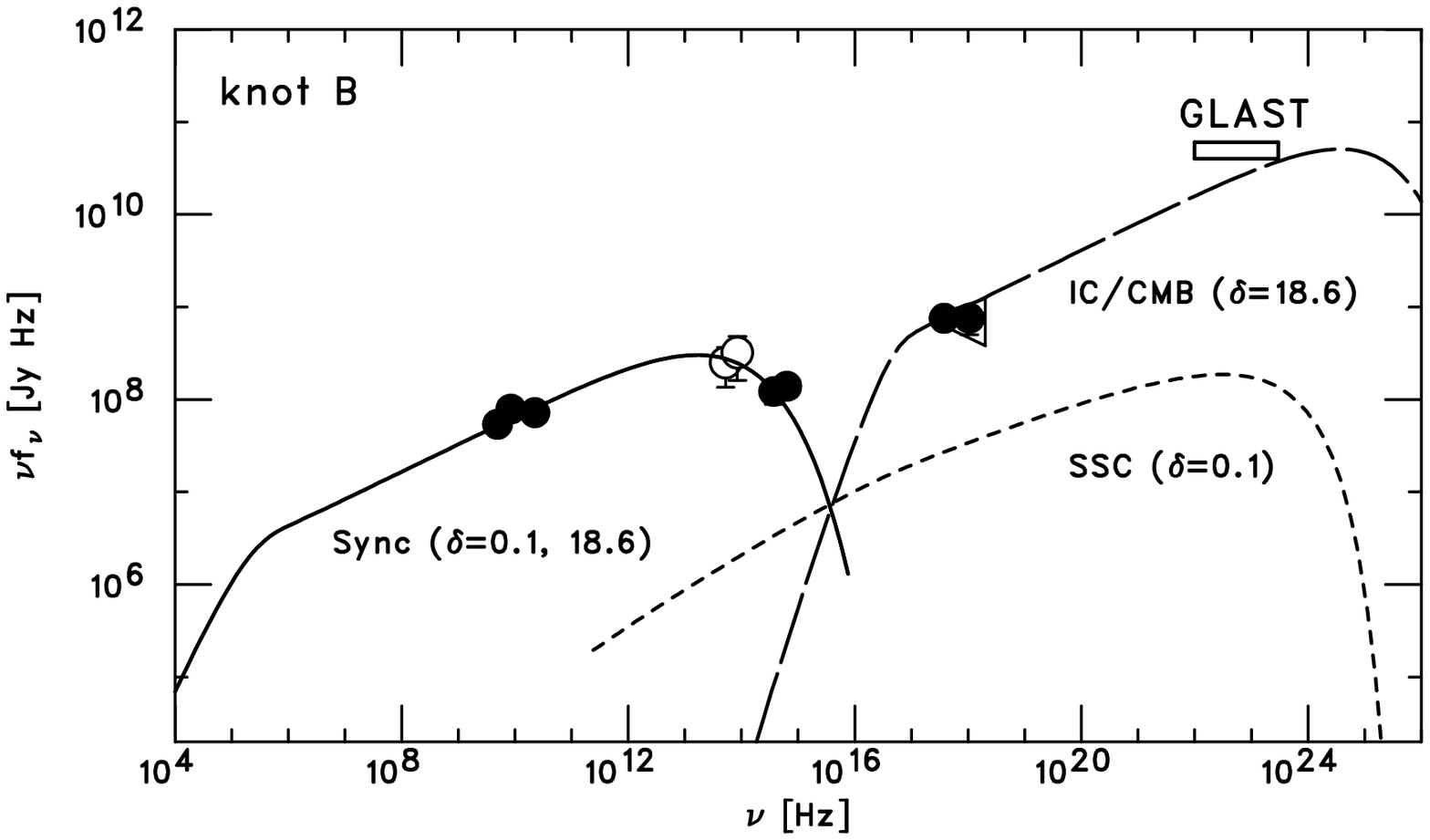}
\plotone{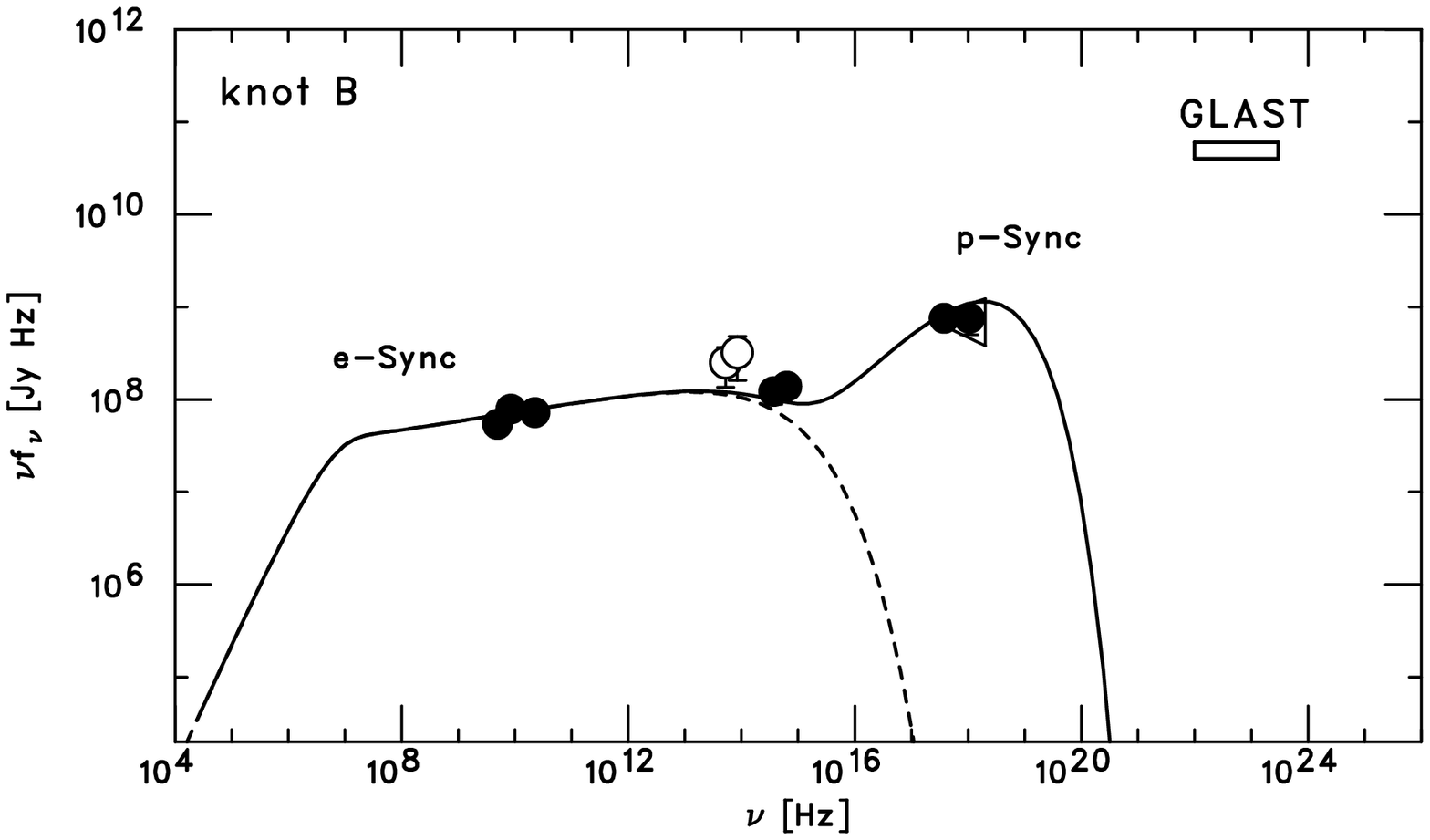}
\caption{({\it Top})
Synchrotron+IC/CMB modeling ({\it solid} and {\it dashed} curves)
for knot B.  The SED points are same as Fig.~\ref{fig:SED}. 
A large beaming factor of  $\delta=18.6$ is required to explain the X-ray flux. 
For illustrative purpose, it is also shown that 
if instead we invoke de-beaming ($\delta =0.1$: a {\it dotted} curve), 
the SSC luminosity  becomes comparable to the synchrotron one.
The expected sensitivity of the GLAST satellite (1 year $5\sigma$)  is marked. 
({\it Bottom})
Proton synchrotron model for the high-energy emission with moderate 
beaming $\Gamma = \delta = 3$ ($\theta \simeq 19\degr$). 
The low-energy emission is ascribed to electron synchrotron radiation.
The {\it dotted} curve shows the electron component alone, while the {\it solid} 
line presents synchrotron radiation produced by both protons and electrons. 
Assuming equipartition between protons and magnetic fields, 
the magnetic field strength is $B_{\rm eq} = 1.7\ \rm mG$. 
The production rate of electrons in terms of the energy content is about 5\% of 
that of protons. 
 \label{fig:ICCMB}}
\end{figure}

Figure~\ref{fig:ICSSC} shows the IC/CMB and SSC 
X-ray fluxes as a function of $\delta$, given the radio flux observed for knot B.
Since the normalization of the electron distribution scales as $k \propto \delta^{-2}$
and the beaming pattern of the IC/CMB emission follows $\delta^{3+s}$
 \citep{Der95,GKM01}, 
the IC/CMB flux scales as $k \delta^{3+s} \propto \delta^{1+s} = \delta^{3.4}$. 
The SSC flux scales as $k^2 B^{1+\alpha}\delta^{3+\alpha} \propto \delta^{-2}$, where 
$\alpha = (s-1)/2 = 0.7$. As such, 
if the jet is heavily beamed, $\delta \gg 1$, the IC/CMB flux is dominant.
It reaches the observed X-ray flux at $\delta =18.6$; the corresponding  SED is 
shown in the top panel of Fig.~\ref{fig:ICCMB}.
The SSC becomes a main component for $\delta < 1.4$. Even if we 
invoke a particle-dominated case of $U_e/U_B=5000$, the SSC model 
requires a significant de-beaming of $\delta = 0.13$ to explain the observed X-rays, 
 which is quite unlikely as 
the on-axis version of such jet emission becomes unacceptably luminous.

The radio-emitting electrons 
 emit GeV $\gamma$-rays through the IC/CMB process (Fig.\ \ref{fig:ICCMB}).
The total radiative 
output is dominated by the multi-GeV $\gamma$-rays and their predicted flux just 
reaches the sensitivity offered by the upcoming GLAST satellite 
\citep[see][for the case of 3C 273]{Geo06}. 
The quasar core is expected to show a similar level of $\gamma$-ray 
emission \citep{Sam06b}; steadiness and hard spectrum are signatures of 
the large-scale emission. 
The observation of  PKS~1136$-$135 (or quasar jets in general) 
with GLAST may be able to test the IC/CMB hypothesis. 
Note that the sub-TeV domains are subject to intergalactic absorption, 
which makes it difficult to investigate the jet emission with future 
ground-based Cherenkov telescopes.

The  Doppler factor of $\delta \simeq 19$ required by the IC/CMB model implies 
an uncomfortably small angle between the jet and the line-of-sight, 
 $\theta < 4\degr$.
\citet{Tav06} also derived 
the maximum angle permitted as $\theta_{\max} =3\fdg 8$ for this jet. 
The small angle would make the jet quite long, $>(77/\sin \theta_{\max})\ \rm kpc =1.1\ \rm Mpc$, and the total source extent would be as large as the largest quasar.
The core-to-lobe flux ratio at 5 GHz, $R$, is an indicator 
of jet orientation.
In the case of 1136$-$135, the core-to-lobe ratio of $R = 0.29$ 
is obtained \citep{Sam04}, 
thus designated as a {\it lobe-dominated} quasar. 
According to \citet{Lau97}, a sample of radio-loud quasars with a median 
value of $R = 0.52$ have an average angle 
to the line-of-sight of $\langle \theta \rangle \sim 30\degr$ with 
simple beaming models. 
Also, the mean angle for steep-spectrum radio quasars is estimated as 
$\langle \theta \rangle \sim 30\degr$ \citep{UP95}.
In this respect, a very small viewing angle $\theta < 4\degr$ for a lobe-dominated 
quasar like PKS~1136$-$135 is not favored
in the light of simple unification schemes of radio-loud quasars and radio galaxies 
\citep{UP95}; 
a similar statement was made by \citet{KS05}. 
However, it should be kept in mind that {\it Chandra}-detected jets 
can be biased toward smaller jet angles, 
which alleviates a problem of the jet angle. 
We need a more direct means to distinguish currently proposed models 
(see \S\ref{sec:discri}).

\subsection{Synchrotron Radiation by Second Electron Population}

The large-scale jet in PKS~1136$-$135 is in many respects similar to 
the well-known jet of quasar 3C~273. 
As \citet{Sam06a} have pointed out, in both jets, 
the radio emission brightens monotonically towards the terminal 
hotspot, while the upstream knots are brighter in X-rays. 
The two-component SEDs are also similar \citep[cf,][]{Uch06}.
In the case of 3C~273, since 
both the radio and optical emission are linearly polarized to 
a similar degree and in the same direction, 
it seems reasonable that the high-energy component, which
contributes more than half the optical flux, should also
arise from the synchrotron process \citep{Uch06}.
Therefore we  discuss the optical-to-X-ray component 
of 1136$-$135 in terms of 
synchrotron radiation produced by a second population of high-energy electrons. 
The suggestion of the presence of a second synchrotron component has been made
in previous studies of other large-scale jets \citep[e.g.,][]{JM04,Har04,Che06}.

Let the energy distribution of \emph{accelerated} electrons be 
 $Q(E_e)dE_e\propto E_e ^{-s_2}dE_e$ for an energy interval of interest. 
The power-law slope of electrons is derived from the radiation slope 
as $s_2 = 2\alpha_2$ [see Eq.~(\ref{eq:fnu})]
in a synchrotron-cooling regime as appropriate for optical and X-ray emission; 
the \emph{cooled} electrons 
responsible for the observed radiation have a steeper number index 
of $s_2 +1$ \citep[see e.g.,][]{Kata03}.
Then the spectral index of the high-energy component, 
$\alpha_2 \simeq 0.5\mbox{--}0.75$ 
corresponds to $s_2\simeq 1.0\mbox{--}1.5$. 
However, within the framework of diffusive shock acceleration, 
the hardest possible electron distribution of acceleration 
would be $Q(E_e)\propto E_e ^{-1.5}$, in other words, 
$s_2 \geq 1.5$ \citep[see][]{MD01}. 
The inferred index $s_2\simeq 1.0\mbox{--}1.5$ would violate 
such a theoretical limit. 
This brings into question the idea that the second electron distribution
is formed through the diffusive shock acceleration or that there
is a second electron distribution at all. 
Note that 
the jet in 3C~273 has 
spectral index of $\alpha_2 \simeq 0.7\mbox{--}0.8$ 
corresponding to $s_2\simeq 1.4\mbox{--}1.6$, which was considered to 
be compatible with the shock acceleration theory \citep{Uch06}.

The second synchrotron component in the PKS~1136$-$135 jet 
may instead be due to turbulent acceleration operating in the shear layers 
\citep{Ost00,SO02}. 
The coexistence of 
two distinct types of acceleration, shock and turbulent acceleration, in 
the knots would naturally give rise to a double-synchrotron spectrum. 
More importantly, unlike the shock acceleration, 
a hard spectrum, $s_2 < 1.5$, can be expected 
to form in the case of turbulent acceleration (i.e., second-order 
Fermi acceleration). 
In fully turbulent shear layers, the time scale of turbulent acceleration 
can be estimated as 
$t_{\rm acc} \sim r_g c v_{\rm A}^{-2}$, where $r_g=E_e/(eB)$ is 
an electron gyroradius and 
$v_{\rm A}$ denotes the Alfv\`en velocity. 
The turbulent acceleration has to compete against synchrotron 
losses, which is presumed to dominate over IC losses, with 
a timescale of 
\begin{equation}
t_{\rm syn}
 = \frac{\gamma_e m_e c^2}{(4/3)c\sigma_{\rm T} U_B \gamma_e^2} , 
\end{equation}
where $\sigma_{\rm T}$ is the Thomson cross section, and 
$U_B = B^2/(8\pi)$ is the energy density of the magnetic field.
By equating $t_{\rm acc} = t_{\rm syn}$---namely, balancing the 
acceleration and synchrotron loss rates---one obtains the maximum 
attainable energy limited by synchrotron losses:
\begin{equation}
E_{e,\max} \sim 6\times 10^{15}\ \frac{v_{\rm A}}{c} 
\frac{1}{\sqrt{B_{-4}}}
\ \rm eV, 
\end{equation}
where $B_{-4} = B/(10^{-4} \rm G)$.
To account for the observed X-rays with synchrotron radiation, 
one needs $E_{e,\max} \ga 30\, (B_{-4}\delta)^{-0.5}\ \rm TeV$
\citep[see][]{Uch06}, thus requiring $v_{\rm A} \ga 5\times 10^{-3} c$.
This condition is reasonable with typical jet parameters.

\subsection{Proton Synchrotron Radiation}

Finally, we argue the possibility that 
the optical-to-X-ray emission may be due to 
synchrotron radiation by very high energy {\it protons} \citep{Aha02}.
This model requires that protons in the jet are somehow accelerated 
to very high energies, $E_{p} \sim 10^{18}\ \rm eV$ or more. 
Also, the magnetic field strength must be of the order $B \sim \rm mG$, 
so that energy equipartition between the relativistic {\it protons} and magnetic fields 
can be (roughly) realized. 
The knots and hotspots in the relativistic jets of 
powerful quasars and radio galaxies are indeed 
one of a few potential sites of cosmic-ray acceleration up to 
$E_{p} \sim 10^{20}\ \rm eV$ \citep{Hil84}.
For example,  the turbulent acceleration in the shear layer 
may be able to accelerate such ultra high energy protons \citep{Ost00}.

The characteristic frequency and cooling time of proton synchrotron radiation 
can be written as 
\begin{equation}
\nu_{\rm p-syn} \simeq 2.1 \times 10^{18} \ 
\left( \frac{B}{\rm mG}\right) 
\left( \frac{E_{p}}{10^{18}\ \rm eV} \right)^2\ 
\rm Hz, 
\end{equation}
and 
\begin{equation}
t_{\rm p-syn} \simeq 1.4 \times 10^{8} \ 
\left( \frac{B}{\rm mG}\right) ^{-2}
\left( \frac{E_{p}}{10^{18}\ \rm eV} \right)^{-1}\ 
\rm yr,
\end{equation}
respectively \citep{Aha02}. 
(Note that the synchrotron cooling time is much shorter than 
the photo-meson cooling time at energies relevant here.)
A total energy content of the magnetic field $B= 1\ \rm mG$ integrated over 
a spherical knot with a radius of $r= 1\ \rm kpc$ amounts to 
$W_B = 4.9 \times 10^{57}\ \rm erg$.
If we assume equipartition between radiating protons and 
magnetic fields, $W_{p} = W_B$, the X-ray luminosity of synchrotron radiation 
by ultra-high-energy protons of $E_{p} \sim 10^{18}\ \rm eV$ 
can be estimated roughly as 
\begin{equation}
L_x \sim \frac{\eta W_p}{t_{\rm p-syn}} 
\sim \eta \frac{4.9\times 10^{57}\ \rm erg}{1.4\times 10^{8}\ \rm yr}
\sim 1.1 \times 10^{42}\ \eta\ \rm erg\ s^{-1}, 
\end{equation}
where $\eta \sim 0.1 $ 
denotes the fraction of kinetic energy of protons responsible for the X-ray emission.
Relativistic beaming is not taken into account in this estimate. 
Roughly speaking, the luminosity scales with $B$ and $\delta$ such that 
$L_x \propto (B\delta)^4 $.
The expected luminosity agrees with typical (apparent) luminosity of 
the jet knots of quasars, $L_x \sim 10^{43}\ \rm erg\ s^{-1}$, 
if  $B \delta \sim 6\ \rm mG$.
Thus, very high energy protons rather than a second population of electrons, may 
explain the second synchrotron component extending from the optical to X-rays
provided that both ultra-high-energy protons of $E_{p} \sim 10^{18}\ \rm eV$ 
and a magnetic field of $B \sim \rm mG$ are present in the jet knots. 
Under the proton-synchrotron hypothesis, 
the low-energy synchrotron emission may be accounted for by 
accompanying electrons. 

To be specific, we modeled the high-energy spectral component of knot B 
with proton synchrotron radiation. 
We adopt mild beaming of $\delta = \Gamma =3$ ($\theta \simeq 20\degr$). 
The low-energy spectral component was simultaneously modeled by 
electron synchrotron radiation with taking account of the effects of significant 
synchrotron cooling.  It is assumed that during $\Delta t = 10^7$ years, 
protons and electrons are injected continuously into the emission volume 
($r=1\ \rm kpc$) with 
the distributions characterized by number index $s_p$ and $s_e$ and by 
the maximum energy $E_{p, \max}$ and $E_{e, \max}$: 
$\dot{Q}(E) = k E^{-s} \exp (-E/E_{\max})$.
In the bottom panel of Fig.~\ref{fig:ICCMB}, we present the broadband SED 
reproduced by the proton synchrotron model with 
the equipartition magnetic field of $B= 1.7\ \rm mG$,
the index of the power-law distribution of \emph{accelerated} particles  
$s_p = s_e = 1.8$, and the maximum energy of 
$E_{p, \max} = 4.7\times 10^{17}\ \rm eV$ and 
$E_{e, \max} = 1\times 10^{11}\ \rm eV$.
The injection rate of electrons, in terms of energy content $\int \dot{Q}(E)dE$, 
is about 5\% of that of protons.  The injected electrons suffer from severe  synchrotron 
cooling. Even radio-emitting electrons have a cooling time of only $\sim 10^4$ years,
which requires {\it in-situ} acceleration of electrons. The infrared-emitting 
electrons have a lifetime of $\sim 10^2$ years, implying that each knot is 
a currently (within $10^2$ years) active site of particle acceleration. 

The energy density of protons $U_p$, 
which is equal to $U_B$, can be calculated by 
$U_p \simeq (\Delta t/V) \int \dot{Q_p}(E) dE\simeq 
7.4\times 10^4\ \rm eV\ cm^{-3}$ with $V=4\pi r^3/3$, since 
synchrotron cooling of protons is not effective. 
The kinetic power of the jet, estimated as 
$L_{\rm jet} \sim \pi r^2 \Gamma^2 \beta c\, (U_p+U_B) $, 
amounts to $L_{\rm jet} \sim 2\times 10^{48}\ \rm erg\ s^{-1}$.
This estimate is not sensitive to the choice of $\Gamma$ as long as 
 the equipartition condition is fulfilled. 
A large power has to be carried by the protons and magnetic fields in the jet 
to produce the X-ray emission via a proton-synchrotron process. 
The black hole mass that powers the jet is estimated to be $4.6\times 10^{8} M_{\sun}$
 \citep{Osh02} using the empirical relation to the width of the H$\beta$ line 
 and the optical continuum luminosity. 
The black hole mass  corresponds to 
the Eddington luminosity of $L_{\rm Edd} \simeq 6\times 10^{46}\ \rm erg\ s^{-1}$.
Therefore, the proton synchrotron model requires the jet to be 
{\it super-Eddington}, $L_{\rm jet} \gg  L_{\rm Edd}$.
This issue would pose a problem for the proton synchrotron model, though 
it may be alleviated by assuming that 
the knots present the locations of {\it power peaks} due to 
modulated activity of the central engine.

\subsection{Possible Discriminator of the Radiation Models: 
Polarization of Optical Light}
\label{sec:discri}

We emphasize here that optical polarimetry can be an effective 
 way of discriminating the radiation models  
responsible for the optical-to-X-ray emission of  the jet in  PKS~1136$-$135, 
and of quasar jets in general. 
In the beamed IC interpretation, if the optical fluxes belong to the IC component, 
the optical and X-ray emission are due to Compton up-scattering 
off the amplified CMB by 
high-energy electrons of 
$\gamma_{e} \sim 3$ (optical) and $\gamma_{e} \sim 100$ (X-ray).
Unlike in the case of synchrotron models, 
the X-rays are expected to be {\it unpolarized} and 
the optical light is nearly unpolarized at most a few percent of polarization 
(Uchiyama \& Coppi, in preparation)\footnote{
Compton up-scattering by a highly relativistic jet in which a population of 
cold electrons with $\gamma_{e} \simeq 1$ (in the jet frame) exists, 
namely   ``bulk Comptonization",  
can yield in principle  a  high degree of polarization 
for a jet angle $\theta \simeq \Gamma^{-1}$ \citep{BS87}. For $\gamma_e$ of a 
few as relevant here, however, the polarization is largely suppressed.}.
Precise polarization measurements in the optical can in principle verify (or discard) the 
beamed IC model. Unfortunately, there have been no useful polarization 
observations of quasar jets with {\it HST} so far. 
Only for 3C 273, early {\it HST} polarimetry of the jet was done but 
with the pre-COSTAR FOC and low significances \citep{Tho93}.
As was recently performed for nearby radio galaxies by \citet{Per06}, 
new, deep polarimetry of quasar jets on large-scales is highly encouraged, 
although the optical emission from 
the jet knots is generally faint, say $< \mu \rm Jy$ \citep{Sam04}.

\section{Summary} 
\label{conclusions}

We have detected the infrared emission from the kiloparsec-scale jet 
in the quasar PKS~1136$-$135 with the {\it Spitzer Space Telescope}.
Using the new \emph{Spitzer} data together with deep \emph{Chandra}, 
\emph{Hubble}, and multi-frequency VLA observations from 
\citet{Sam06a}, we construct the broadband spectral energy distributions 
along the large-scale jet. 
The SEDs of the jet knots are comprised of 
two components: radio-to-infrared synchrotron emission and a 
separate high-energy component responsible for X-rays, which may 
extend down to the optical. 
In total, each component has a similar apparent luminosity of the order of 
$L_{\rm rad} \sim  10^{44}\ \rm erg\ s^{-1}$.

The origin of the high-energy component has significant implications to 
the properties of particle acceleration in a relativistic jet as well as  jet dynamics.
We consider three radiation models that have been proposed 
to explain the high-energy component: 
inverse-Compton scattering in a highly relativistic jet, 
synchrotron radiation by a second electron population, and 
synchrotron radiation by highly energetic protons.

In terms of the double synchrotron scenario, 
a flat spectrum of the second synchrotron component, $\alpha_2 \sim 0.5$, would 
turn down diffusive shock acceleration as the mechanism of electron acceleration 
to very high energies,  $> 30$ TeV.
Instead turbulent acceleration in the shear layers may explain the second population 
of electrons giving rise to the flat spectrum. 
In the beamed IC interpretation, a jet orientation with respect to the observer 
of $\theta < 4\degr$ is required, which is not quite favorable given the small 
core-to-lobe ratio of this object. Future optical polarimetry 
or $\gamma$-ray observations with GLAST may be able to test the beamed 
IC model. 
The proton synchrotron model requires a relatively 
large magnetic field of $\sim 1.7$ mG, 
and consequently large kinetic power of 
$L_{\rm jet} \simeq 2 \times 10^{48}\ {\rm erg}\ {\rm s}^{-1} \sim 30\, L_{\rm Edd}$, with efficient acceleration of protons to  $E_{p, \max} \simeq 5\times 10^{17}\ \rm eV$.
If this is the case, one can probe particle acceleration at very high energies, 
higher than any other known objects in the Universe.


\acknowledgments

We thank the referee, Eric Perlman, for his careful reading of the 
manuscript and valuable suggestions. 
This work is based on observations made with the Spitzer Space Telescope,
which is operated by the Jet Propulsion Laboratory, California Institute of 
Technology under NASA contract 1407. Support for this work was provided 
by NASA through Contract Number RSA 1265389 issued by JPL/Caltech.
The National Radio Astronomy Observatory is operated by Associated
Universities, Inc.\ under a cooperative agreement with the National Science
Foundation.






\end{document}